\documentclass{PoS}

\usepackage{amsmath}
\usepackage{textcomp}

\title{The General Antiparticle Spectrometer (GAPS) - Hunt for dark matter using low-energy antideuterons}

\ShortTitle{The General Antiparticle Spectrometer (GAPS)}

\author{\speaker{Ph. von Doetinchem}, St.Boggs\\
University of California at Berkeley, Space Sciences Laboratory\\
\email{doetinchem@ssl.berkeley.edu}}
\author{T. Aramaki, Ch. Hailey, J. Koglin, N. Madden, K. Mori\\
Columbia University, Astronomy and Astrophysics}
\author{F. Gahbauer\\
Columbia University, Astronomy and Astrophysics\\
University of Latvia, Atomic and Molecular Physics Laboratory}
\author{H. Fuke, T. Yoshida\\
Institute of Space and Astronautical Science, Japan Aerospace Exploration Agency}
\author{W. Craig\\
Lawrence Livermore National Laboratory}
\author{I. Mognet, R. Ong, T. Zhang, J. Zweerink\\
University of California at Los Angeles, Physics and Astronomy}

\abstract{The GAPS experiment is foreseen to carry out a dark matter search using a novel detection approach to detect low-energy cosmic-ray antideuterons. The theoretically predicted antideuteron flux resulting from secondary interactions of primary cosmic rays with the interstellar medium is very low. So far not a single cosmic antideuteron has been detected by any experiment, but well-motivated theories beyond the standard model of particle physics, e.g., supersymmetry or universal extra dimensions, contain viable dark matter candidates, which could led to a significant enhancement of the antideuteron flux due to self-annihilation of the dark matter particles. This flux contribution is believed to be especially large at small energies, which leads to a high discovery potential for GAPS. 

GAPS is designed to achieve its goals via a series of ultra-long duration balloon flights at high altitude in Antarctica, starting in 2014. The detector itself will consist of 13 planes of Si(Li) solid state detectors and a time of flight system. The low-energy antideuterons ($<0.3$\,GeV/n) will be slowed down in the Si(Li) material, replace a shell electron, and form an excited exotic atom. The atom will be deexcited by characteristic x-ray transitions and will end its life by forming an annihilation pion star. This unique event structure will allow for nearly background free detection. To prove the performance of the different detector components at stratospheric altitudes, a prototype flight will be conducted in 2011 from Taiki, Japan.}

\FullConference{Identification of Dark Matter 2010-IDM2010\\
		July 26-30, 2010\\
		Montpellier France}

\begin{document}

\section{Introduction}

Measurements of cosmic rays and gamma rays have a long tradition of delivering interesting insights into several areas of physics, e.g., particle physics, astrophysical objects, and the interstellar medium. Nowadays models are available that are in good agreement with the measured fluxes \cite{galprop}. Nevertheless, cosmic rays remain an interesting field to study new phenomena like dark matter, as yet unknown astrophysical objects, or even baryogenesis.

\subsection{Cosmic rays and dark matter} 

This work will concentrate on the dark matter aspect. Even if the existence of dark matter seems to be proven \cite{darkmatter}, an exciting question remains concerning its nature. This is particularly interesting as the latest results of the positron flux measurements (PAMELA \cite{pamela}) and the combined positron and electron flux measurements (ATIC \cite{atic} and Fermi LAT \cite{fermi}) may be utilized to constrain dark matter properties. 

The general idea behind these interpretations is briefly explained in the following: It was shown that dark matter particle candidates do not exist within the standard model of particle physics, but well-motivated theories beyond this model, like supersymmetry \cite{neutralino} and universal extra dimensions \cite{lkp}, contain viable candidates. It is now believed that these particles are Majorana particles and are therefore able to self-annihilate. Whatever the exact physical processes in the underlying theories for these annihilation processes, it is assumed that standard model particles are among the final products, which then in turn contribute to the total flux of cosmic rays. The actual shape of these additional contributions are then used to constrain the parameters of the dark matter model under study. As the calculated dark matter annihilation fluxes are generally small, it appears to be very difficult or even impossible to look for deviations in the most abundant cosmic-ray species like protons and alphas. Therefore people concentrate on electrons, positrons, and antiprotons, which have (much) smaller fluxes compared to the primary species. The PAMELA, ATIC, and Fermi LAT positron and electron results show deviations from the predicted fluxes calculated within the frameworks of the available cosmic-ray models. These deviations could be interpreted as additional fluxes coming from dark matter annihilations.

Despite the success of the aforementioned experiments, a few drawbacks exist for a reliable dark matter interpretation using positrons and electrons: The observed deviations are relatively small and might be also interpreted in more standard astronomical ways. Furthermore, the predicted dark matter annihilation fluxes tend to be even smaller than the fluxes needed to explain the deviations and often require the introduction of some kind of boosting mechanism \cite{boost}. This makes it even harder to disentangle the different contributions.

\subsection{Antideuterons as a signature for dark matter} 

\begin{figure}
\centerline{\includegraphics[width=0.93\linewidth]{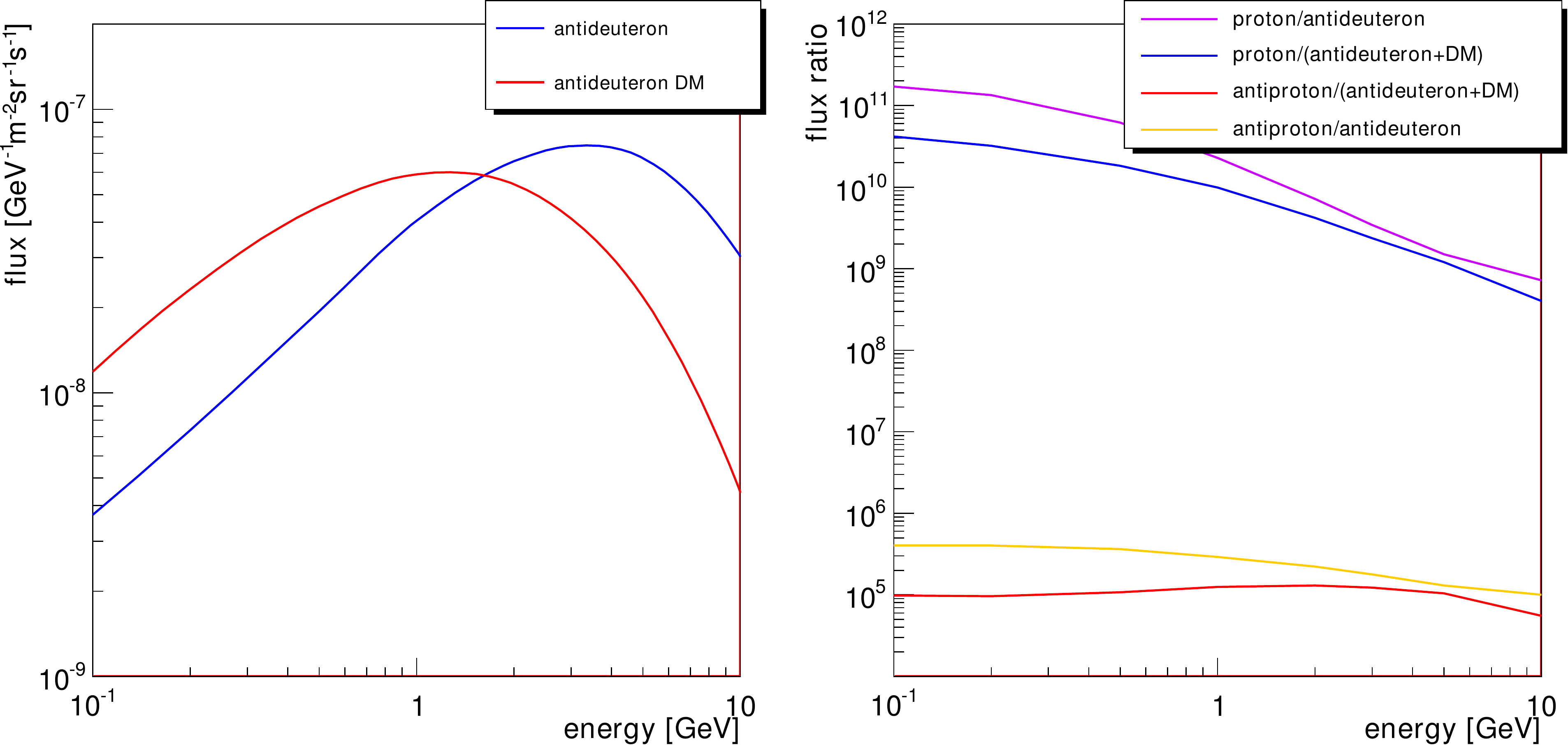}}\caption{\label{f-dbar_flux}\textbf{\textit{Left)}} Antideuteron fluxes from secondary interactions of cosmic rays (blue) with the interstellar medium and from dark matter annihilations (red) \cite{antideuteron,antideuteroncui}. \textbf{\textit{Right)}} Ratio of (anti)proton fluxes to antideuteron fluxes with and without an extra contribution from dark matter annihilations \cite{galprop,antideuteron,antideuteroncui}.}
\end{figure}

A promising channel in the field of indirect dark matter detection is the measurement of antideuterons in cosmic rays \cite{antideuteron}. As cosmic antideuterons have never been measured, only calculations exist. The left hand side of Fig.~\ref{f-dbar_flux} shows the predicted antideuteron flux from interactions of primary cosmic rays with the interstellar medium  \cite{antideuteron} and the flux from dark matter annihilation in a generic model  \cite{antideuteroncui}. It is interesting to note that the dark-matter-induced flux exceeds the secondary flux below 1-2\,GeV without using any boosting mechanism. What makes the antideuteron detection now very challenging becomes immediately evident by looking at the ratios of (anti)proton to antideuteron fluxes (Fig.~\ref{f-dbar_flux}, right). Below 1\,GeV, protons (antiprotons) are about $10^{10}$ to $10^{11}$ ($10^5$ to $10^6$) more abundant than antideuterons. Therefore any attempt to measure reliably cosmic antideuterons needs an exceptionally strong particle identification.

\section{The GAPS experiment}

The General Antiparticle Spectrometer (GAPS) is designed to measure low-energy cosmic antideuterons and must therefore fly at an attitude with small geomagnetic cut-off \cite{gaps}. It is planned to carry out a series of (ultra-)long duration balloon flights from the South Pole starting in 2014.

\begin{figure}
\centerline{\includegraphics[width=0.93\linewidth]{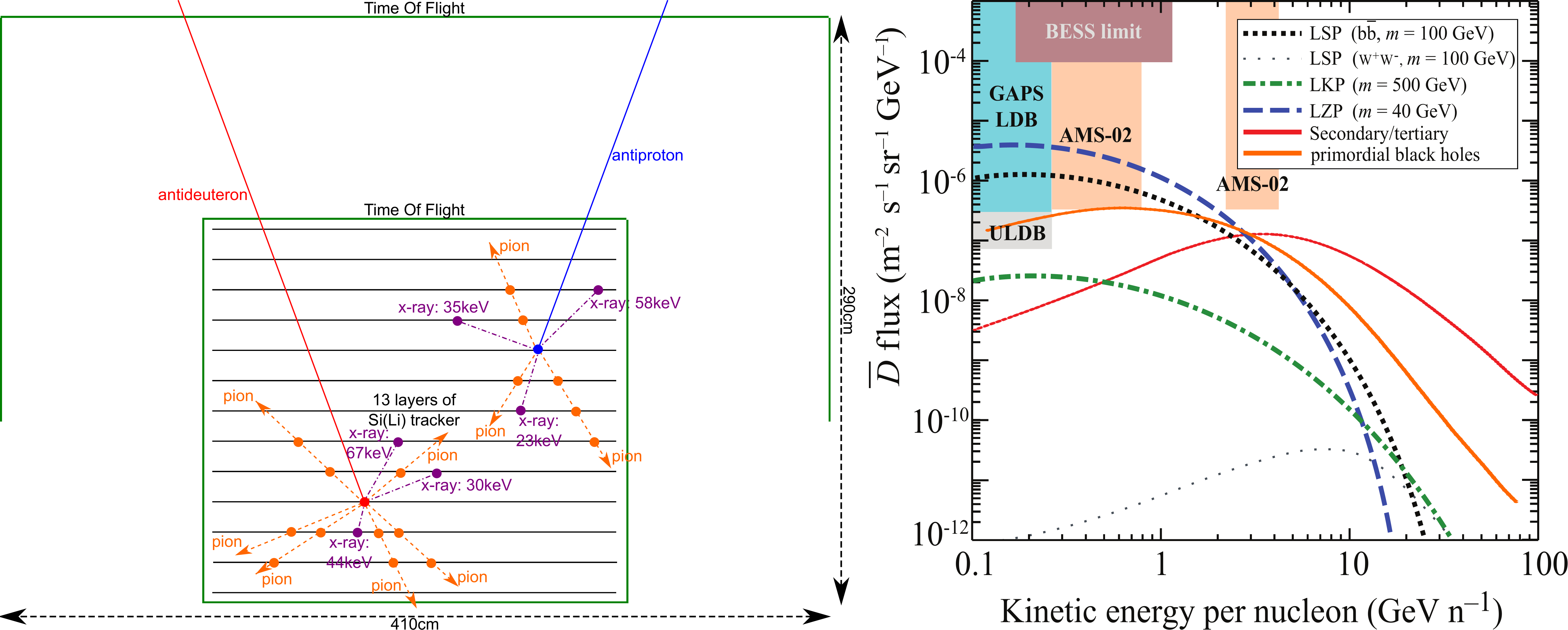}}\caption{\label{f-bgaps_background_sensitivity}\textbf{\textit{Left)}} GAPS detector concept with antiproton and antideuteron signatures. \textbf{\textit{Right)}} Antideuteron sensitivity for GAPS and AMS-02 (superconducting magnet setup) with fluxes from different models \cite{gaps}.}
\end{figure}

\subsection{Detector}

As the predicted antideuteron flux is small, GAPS is designed to have a large geometrical acceptance of about 2.7\,m$^2$sr (Fig.~\ref{f-bgaps_background_sensitivity}, left). The core of the detector is a tracker with 13 layers made of 270 circular, Lithium-doped Silicon modules, which have an escape fraction for x-rays around 20\,keV, a high energy resolution of 3\,keV and a timing resolution of 50\,ns. Each module is 2-3\,mm thick and segmented in eight strips with an active area of about 8\,cm$^2$. The electronics will be designed to resolve both x-rays in the range of 10 to 100\,keV and charged particles with energy depositions up to 200\,MeV. The tracker will be enclosed in a box of plastic scintillators, which will be surrounded by another half-cube of plastic scintillators. This system of plastic scintillators forms the time of flight system (TOF) and is needed to track charged particles and to measure their velocities. For a good timing resolution the readout is done by photomultiplier tubes.

\subsection{Antideuteron identification}

In comparison to the more standard identification techniques implemented by BESS \cite{bessantideuteron} and AMS-02 \cite{amsantideuteron} using magnetic deflection, GAPS utilizes a novel detection approach to clearly identify low-energy antideuterons. The idea is to stop low-energy antideuterons in the tracker material, to replace a shell electron of the target with this antideuteron, and to form an excited exotic atom. A capture of an antideuteron is very likely if its kinetic energy is in the range of the ionization energy. At the beginning this exotic atom will be in a high angular momentum state and will ionize the bound electrons quickly (Auger process). After the complete depletion of shell electrons, a Hydrogen-like exotic atom with an antideuteron surrounding the nucleus is left over and characteristic x-ray ladder transitions will deexcite the atom further, starting from level $n\approx19$. In aa first order approximation, the transition energies can be calculated using the Rydberg formula respecting the reduced mass of the Silicon nucleus-antideuteron system. At the end of the ladder transitions the antideuteron will annihilate in a hadronic interaction with the nucleus and produce pions and protons.

The detector described in the last section is able to measure the velocity and the charge of the incoming particle in the TOF as well as the stopping depth of a particle in the tracker, and it can resolve the characteristic x-ray energies and track the pions and protons.

\subsection{Backgrounds and sensitivity}

The main source of background for the antideuteron signal comes from antiprotons. Typical signatures for antideuterons and antiprotons at the same velocity are compared in Fig.~\ref{f-bgaps_background_sensitivity} on the left hand side. It becomes obvious that a good depth sensing and x-ray energy resolution along with a reliable tracking and counting of pions are essential for a high background rejection. Other sources for backgrounds are cosmic rays like protons and electrons in coincidence with x-rays, which might be able to fake an antideuteron event. Furthermore, the irreducible antideuteron flux from interactions of cosmic rays with the atmosphere needs to be considered and should roughly be on the same order as the secondary cosmic background.

The right hand side of Fig.~\ref{f-bgaps_background_sensitivity} shows the sensitivity for GAPS assuming a (ultra-)long duration balloon flight of 60 (300) days. In addition, the figure shows different scenarios, which give reasonable fluxes within the GAPS sensitivity.

\section{Prototype experiment}

\begin{figure}
\centerline{\includegraphics[width=0.75\linewidth]{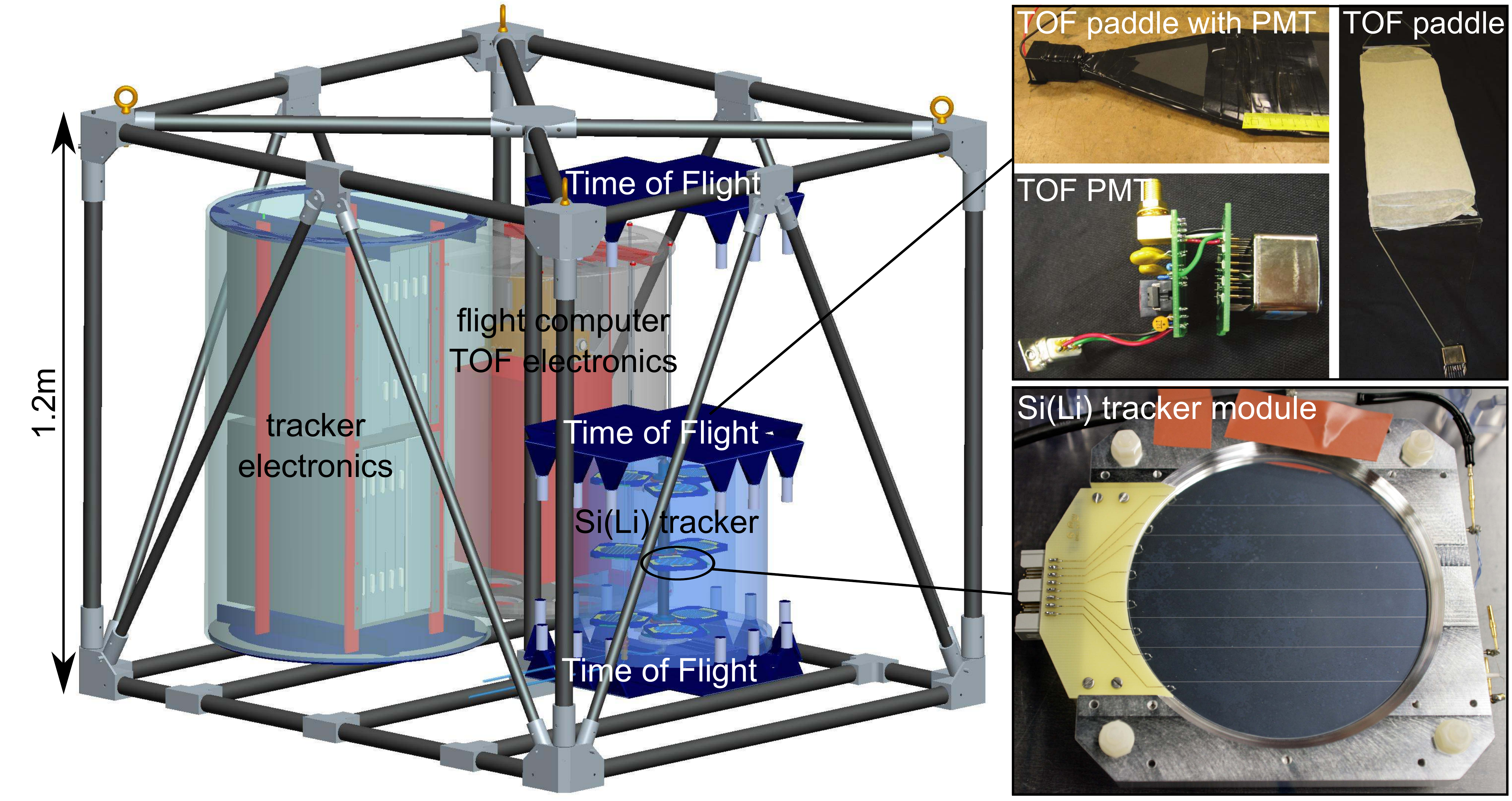}}\caption{\label{f-pGAPS_overview} Prototype GAPS experiment with detailed views of the time of flight system and the Si(Li) tracker modules.}
\end{figure}

A GAPS prototype known as pGAPS (Fig.~\ref{f-pGAPS_overview}) is scheduled for a balloon flight from Taiki, Japan for the late summer of 2011. The flight is expected to last a few hours at an altitude of 33\,km. The main goal of this flight is to demonstrate a stable and low noise operation of the detector components at float altitude and ambient pressure. It is essential for the final GAPS instrument to demonstrate the Si(Li) cooling approach and to verify the thermal model. At the same time this flight will be used to study the incoherent background level in a flight-like configuration. The structure of the balloon gondola frame will be a cube with an edge length of 1.2\,m. The total weight of the detector will be about 450\,kg, and the power consumption about 450\,W.

\subsection{Si(Li) tracker}

The tracker (Fig.~\ref{f-pGAPS_overview}, lower right) will be composed of nine commercially available Si(Li) modules manufactured by Semikon Detektor GmbH of J\"{u}lich, Germany, which will be arranged in three planes. The distance between the individual planes will be 20\,cm. Each module is 2.5\,mm thick, has a circular active area with a diameter of 9.4\,cm, and is divided into eight strips. The strips are on the p+ side and are contacted by implanted Boron. The n+ side has Lithium contacts. In order to achieve a good x-ray detection, the goal is to have an energy resolution of 3\,keV for 60\,keV x-rays at an ambient pressure of 8\,mbar and a temperature of -35\textdegree C.

\subsection{Time of flight system}

The time of flight system (Fig.~\ref{f-pGAPS_overview}, upper right) consists again of three planes. Two planes are above the tracker and one is below. The spacing between the top and bottom layer is 1\,m. Each plane consists of $3\times3$ crossed individual paddles. The paddles are made of Bicron BC-408 material and are 50\,cm long, 15\,cm wide, and 3\,mm thick. Each end will be read out by Hamamatsu R-7600 photomultiplier tubes. The time resolution of 500\,ps translates into a proton velocity resolution of about 6.5\,\% at 300\,MeV kinetic energy.

\subsection{Readout}

The tracker will have a dual readout for each of the 72 channels to be able to resolve the energy depositions of charged particles as well as x-rays. These readout electronics will be placed in a watertight vessel and operated at ambient pressure. The TOF will be read out by a VME system, which is placed in a watertight pressure vessel. The same vessel will also contain the PC-104 flight computer, which will be the interface during flight for all commanding and data handling. 

The TOF will be responsible to generate the main instrument trigger during flight, but it will also be possible to run the tracker in self-trigger mode for calibrations with a x-ray tube.

\subsection{Simulation}

\begin{figure}
\centerline{\includegraphics[width=0.93\linewidth]{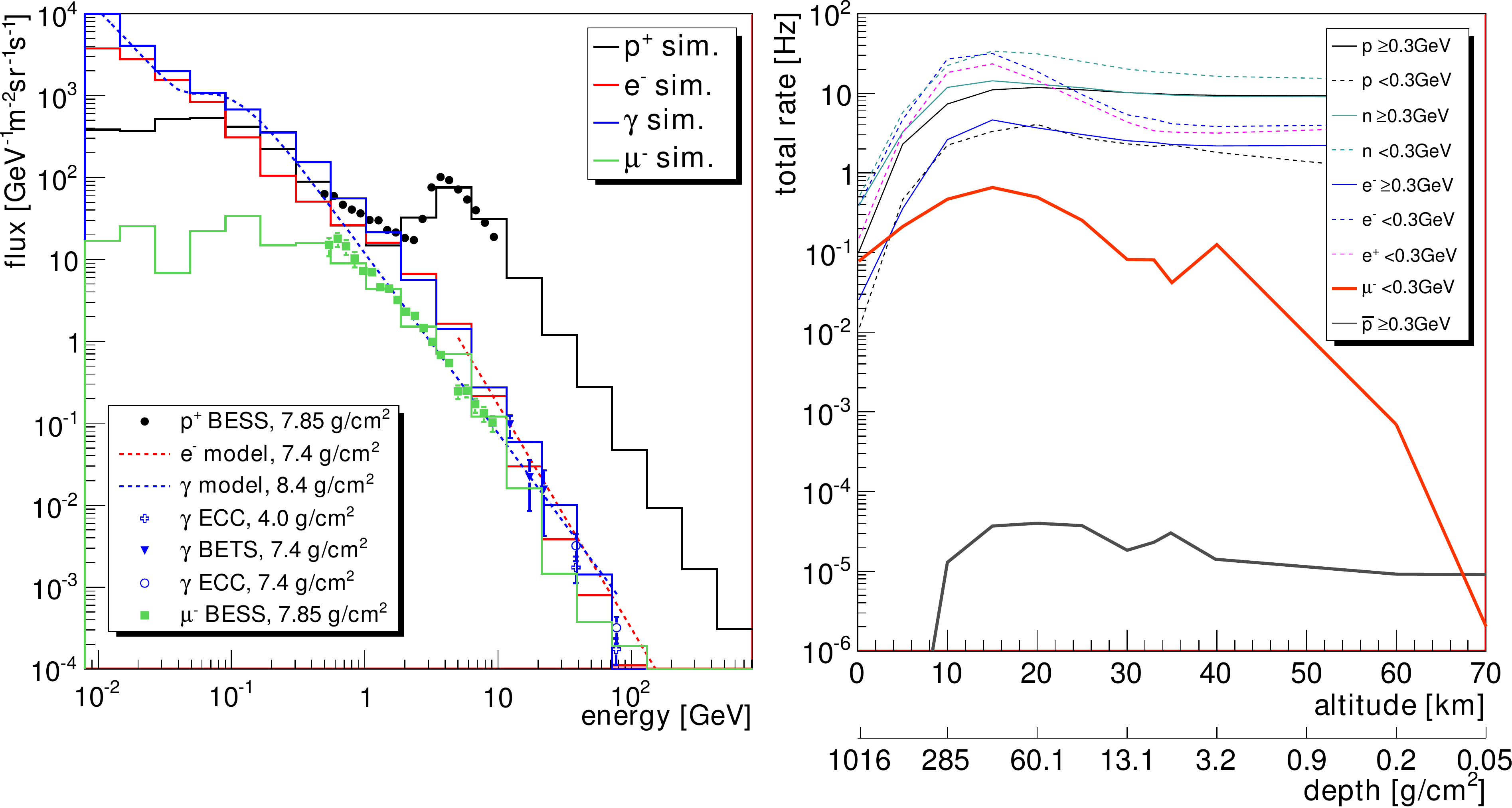}}
\caption{\label{f-all_33km}\textbf{\textit{Left)}} Comparison between available data and models with Planetocosmics simulations in Taiki at 33\,km altitude \cite{bess,nishimura,mizuno}. \textbf{\textit{Right)}} Highest particle rates for pGAPS respecting the angular acceptance (total acceptance: 0.054\,m$^2$sr) as a function of altitude in Taiki for energies below and above 0.3\,GeV.}
\end{figure}

Reliable simulations of the pGAPS instrument play a key role in the development process of the full GAPS instrument because a deep understanding of the possible background signatures is necessary. Therefore, simulations with GEANT \cite{geant} are under development already for the pGAPS detector to study the expected detector responses. 

Additionally, it is important to study the influence of the atmosphere, which alters the composition of cosmic rays through interactions (grammage of atmospheric matter above 33\,km: $\approx8.4$\,g/cm$^2$). Simulations were carried out using a modified Planetocosmics package (based on GEANT) to study both the atmospheric influence and the geomagnetic effects \cite{planeto}. A comparison of these simulations with available data and models is shown in Fig.~\ref{f-all_33km} on the left hand side. As data and simulations are in good agreement, the simulations are used to calculate the expected fluxes in Taiki (N 42\textdegree, E 143\textdegree) and shown in Fig.~\ref{f-all_33km} on the right hand side. At flight altitude the strongest particle backgrounds are neutrons, protons, electrons, and positrons. The total rate for a TOF trigger signal generated by all three planes is about 30\,Hz. It is hereby interesting to note that the low-energy charged particle fluxes are of atmospheric origin because of a high geomagnetic cut-off rigidity at Taiki of about 8\,GV. These particle fluxes are going to be fed into the detector simulation to carry out detailed studies of the detector behavior and reconstruction. The final goal is to compare the simulations to the flight data and to realistically adapt the underlying simulation models for an extremely reliable detector and background simulation of the final instrument.

\section{Conclusion and outlook}

The measurement of the low-energy antideuteron flux is a promising way to search for dark matter indirectly. The GAPS experiment is specifically designed to perform this task by stopping antideuterons and forming and detecting exotic atoms. It is planned to carry out (ultra-)long duration balloon flights from the South Pole starting from 2014. A prototype experiment is currently under construction, and a flight is scheduled for the summer of 2011 from Taiki, Japan. The hardware, software, and simulations for this flight are currently under development.

\end{document}